\documentclass{article}
\usepackage{amsfonts}
\usepackage{spconf, amsmath, graphicx, epsfig}
\usepackage{amsmath, amsthm, amssymb, epsfig}
\usepackage{xcolor}
\usepackage{hyperref}
\usepackage{algorithm}
\title{Optimizing Quantum Federated Learning Based on Federated Quantum Natural Gradient Descent}
\name{Jun Qi$^{1, 2}$, Xiao-Lei Zhang$^{3}$, Javier Tejedor$^4$}
\address{
$^1$ Electronic Engineering, Fudan University, Shanghai, China \\
$^2$ Electrical and Computer Engineering, Georgia Institute of Technology, Atlanta, GA, USA \\
$^3$ Marine Science and Technology, Northwestern Polytechnical University, Xian, China	\\
$^4$ Institute of Technology, Universidad San Pablo-CEU, CEU Universities, Urb. Montepríncipe, Madrid, Spain \\
}

\begin{document}
\ninept
\maketitle
\begin{abstract}
Quantum federated learning (QFL) is a quantum extension of the classical federated learning model across multiple local quantum devices. An efficient optimization algorithm is always expected to minimize the communication overhead among different quantum participants. In this work, we propose an efficient optimization algorithm, namely federated quantum natural gradient descent (FQNGD), and further, apply it to a QFL framework that is composed of a variational quantum circuit (VQC)-based quantum neural networks (QNN). Compared with stochastic gradient descent methods like Adam and Adagrad, the FQNGD algorithm admits much fewer training iterations for the QFL to get converged. Moreover, it can significantly reduce the total communication overhead among local quantum devices. Our experiments on a handwritten digit classification dataset justify the effectiveness of the FQNGD for the QFL framework in terms of a faster convergence rate on the training set and higher accuracy on the test set. 
\end{abstract}

\section{Introduction}

Deep learning (DL) technologies have been successfully applied in many machine learning tasks such as speech recognition (ASR)~\cite{huang2014historical}, natural language processing (NLP)~\cite{hirschberg2015advances}, and computer vision~\cite{voulodimos2018deep}. The bedrock of DL applications highly relies on the hardware breakthrough of the graphic processing unit (GPU) and the availability of a large amount of training data~\cite{qi2020analyzing, qi2019theory}. However, the advantages of large-size DL models, such as GPT-3~\cite{brown2020language} and BERT~\cite{devlin2018bert}, are faithfully attributed to the significantly powerful computing capabilities that are only privileged to big companies equipped with numerous costly and industrial-level GPUs. With the rapid development of noisy intermediate-scale quantum (NISQ) devices~\cite{Preskill2018quantumcomputingin, egan2021fault, guo2021testing}, the quantum computing hardware is expected to speed up the classical DL algorithms by creating novel quantum machine learning (QML) approaches like quantum neural networks (QNN)~\cite{cerezo2021variational, power_data, du2021learnability, huang2022quantum} and quantum kernel learning (QKL)~\cite{huang2022quantum}. The VQC-based QNN seeks to parameterize a distribution through some set of adjustable model parameters, and the QKL methods utilize quantum computers to define a feature map that projects classical data into the quantum Hilbert space. Both QML methods have advantages and disadvantages in dealing with different machine learning tasks and it could not be simply claimed which one is the most suitable choice. However, two obstacles prevent the NISQ devices from applying to QML in practice. The first challenge is that the classical DL models cannot be deployed on NISQ devices without model conversion to quantum tensor formats~\cite{qi2021qtn, chen2021end}. For the second challenge, the NISQ devices admit a few physical qubits such that insufficient qubits could be spared for the quantum error correction~\cite{ball2021real, egan2021fault, guo2021testing}. More significantly, the representation power of QML is quite limited to the small number of currently available qubits~\cite{qi2022theoretical} and the increase of qubits may lead to the problem of Barren Plateaus~\cite{mcclean2018barren}.

To deal with the first challenge, in this work, we introduce a variational quantum algorithm, namely a variational quantum circuit (VQC), to enable QNN to be simulated on the currently available NISQ devices. The VQC-based QNNs have attained even exponential advantages over the DL counterparts on exclusively many tasks like ASR~\cite{yang2021decentralizing, qi2021classical}, NLP~\cite{yang2022bert}, and reinforcement learning~\cite{chen2020variational}. As for the second challenge, distributed QML systems, which consist of local quantum machines, can be set up to enhance the quantum computing power. One particular distributed QML architecture is called quantum federated learning (QFL), which aims to build a decentralized computing model derived from a classical FL~\cite{li2020federated}.

\begin{figure}[htbp]
\centerline{\epsfig{figure=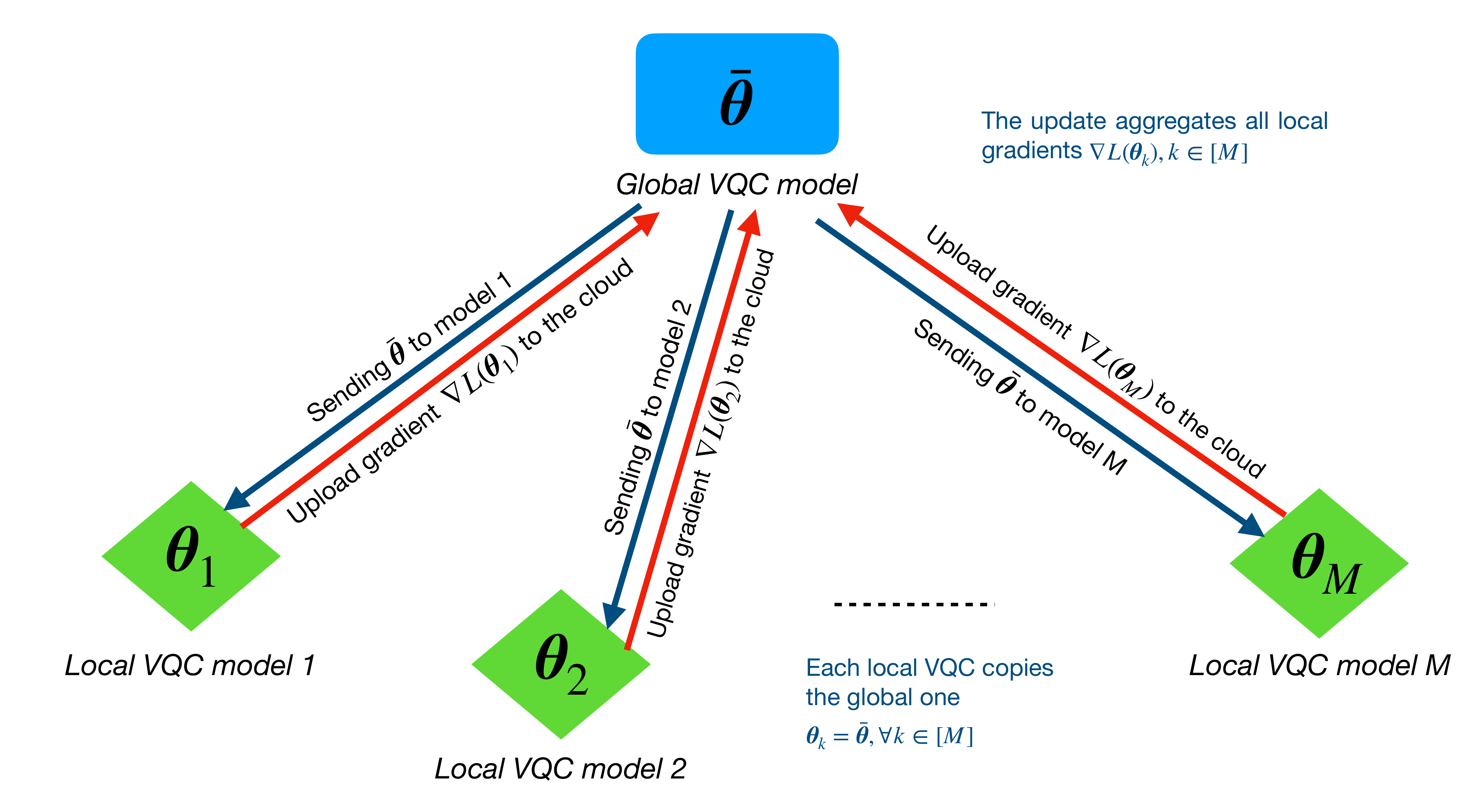, width=85mm}}
\caption{{\it An illustration of quantum federated learning. The global VQC parameter $\bar{\boldsymbol{\theta}}$ is first transmitted to local VQCs $\boldsymbol{\theta}_{k}$. Then, the updated gradients $\nabla\mathcal{L}(\boldsymbol{\theta}_{k})$ based on the participants' local data are sent back to the centralized server and then they are aggregated to update the parameters of the global VQC.}}
\label{fig:qfl}
\end{figure}

Kone{\v{c}}n{\`y} \emph{et al.}~\cite{konevcny2016federated} first proposed the FL strategies to improve the communication efficiency of a distributed computing system, and McMahan \emph{et al.}~\cite{mcmahan2017communication} set up the FL systems with the concerns in the use of big data and a large-scale cloud-based DL~\cite{shokri2015privacy}. The FL framework depends on the advances in hardware progress, making tiny DL systems practically powerful. For example, an ASR system on the cloud can transmit a global acoustic model to a user's cell phone and then send the updated information back to the cloud without collecting the user's private data on the centralized computing server. As shown in Figure~\ref{fig:qfl}, the QFL system is similar to a classical FL system and differs from distributed learning in several ways as follows: (a) the datasets in the framework of QFL are not necessarily balanced; (b) the data in QFL are not assumed to be generated from an independent and identical (i.i.d.) distribution. 

Chen \emph{et al.}~\cite{chen2021federated} demonstrates the QFL architecture that is built upon the classical FL paradigm, where the central node holds a global VQC and receives the trained VQC parameters from participants' local quantum devices. Therefore, the QFL model, which inherits the advantages of the FL framework, can unite tiny local quantum devices to generate a powerful global one. This methodology helps to build a privacy-preserving QML system and leverages quantum computing to further boost the computing power of the classical FL. As shown in Figure~\ref{fig:qfl}, our proposed QFL and FL differ in the models utilized in federated learning systems, where QFL employs VQC models instead of their classical DL counterparts for FL. More specifically, the QFL comprises a global VQC model deployed on the cloud, and there are $M$ local VQC models assigned to users' devices. The training process of QFL involves three key procedures: (1) the parameters of global VQC model $\bar{\boldsymbol{\theta}}$ are transmitted to $K$ local participants' devices; (2) each local VQC first adaptively trains its own model based on the local users' data, and then separately sends the model gradients $\nabla\mathcal{L}(\boldsymbol{\theta}_{k})$ back to the centralized platform; (3) the uploaded gradients from local participants are averagely aggregated to create a global gradient to update further the global model parameters $\bar{\boldsymbol{\theta}}$. 

Despite the advantages of QFL in practice, an inherent bottleneck of QFL is the communication overhead among different VQC models, which bounds up with the performance of QFL. To reduce the cost of communication overhead, we expect a more efficient training algorithm to speed up the convergence rate such that fewer counts of global model updates can be attained. Based on the above analysis, in this work, we put forth a federated quantum learning algorithm, namely federated quantum natural gradient descent (FQNGD), for the training of QFL. The FQNGD algorithm, developed from the quantum natural gradient descent (QNGD) algorithm, admits a more efficient training process for a single VQC~\cite{stokes2020quantum}. In particular, Stokes \emph{et al.}~\cite{stokes2020quantum} first claimed that the Fubini-Study metric tensor could be employed for the QNGD. Besides, compared with the work~\cite{chen2021federated}, the gradients of VQC are uploaded to a global model rather than the VQC parameters of local devices such that the updated gradients can be collected without being accessed to the VQC parameters as shown in~\cite{chen2021federated}.

\section{Variational Quantum Circuit}

\begin{figure}
\centerline{\epsfig{figure=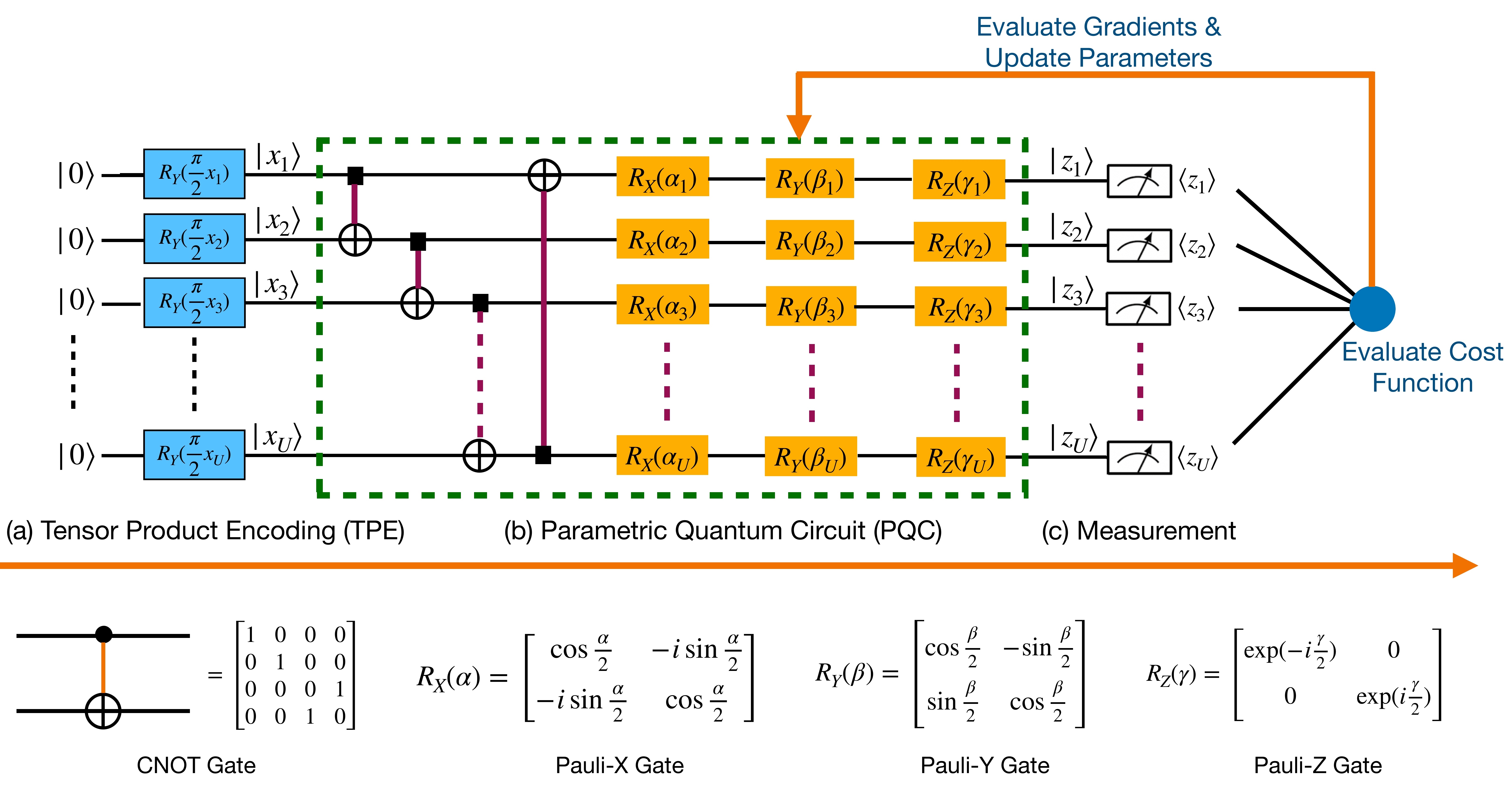, width=90mm}}
\caption{{\it The VQC is composed of three components: (a) TPE; (b) PQC; (c) Measurement. The TPE utilizes a series of $R_{Y}(\frac{\pi}{2} x_{i})$ to transform classical inputs into quantum states. The PQC consists of CNOT gates and single-qubit rotation gates $R_{X}$, $R_{Y}$, $R_{Z}$ with trainable parameters $\boldsymbol{\alpha}$, $\boldsymbol{\beta}$, and $\boldsymbol{\gamma}$. The CNOT gates are non-parametric and impose the property of quantum entanglement among qubits, and $R_{X}$, $R_{Y}$ and $R_{Z}$ are parametric gates and can be adjustable during the training stage. The PQC model in the green dash square is repeatably copied to build a deep model. The measurement converts the quantum states $\vert z_{1}\rangle, \vert z_{2}\rangle, ..., \vert z_{U}\rangle$ into the corresponding expectation values $\langle {z}_{1} \rangle, \langle {z}_{2} \rangle, ..., \langle {z}_{U} \rangle$. The outputs $\langle {z}_{1} \rangle, \langle {z}_{2} \rangle, ..., \langle {z}_{U} \rangle$ is connected to a loss function and the gradient descent algorithms can be used to update the VQC model parameters. Besides, both CNOT gates and $R_{X}$, $R_{Y}$ and $R_{Z}$ correspond to unitary matrices as shown below the VQC framework.}}
\label{fig:vqc}
\end{figure}

An illustration of VQC is shown in Figure~\ref{fig:vqc}, where the VQC model consists of three components: (a) tensor product encoding (TPE); (b) parametric quantum circuit (PQC); (c) measurement. The TPE initializes the input quantum states $\lvert x_{1} \rangle$, $\lvert x_{2} \rangle$, ..., $\lvert x_{U} \rangle$ from the classical inputs $x_{1}$, $x_{2}$, ..., $x_{U}$, and the PQC operator transforms the quantum states $\lvert x_{1} \rangle$, $\lvert x_{2} \rangle$, ..., $\lvert x_{U} \rangle$ into the output quantum states $\lvert z_{1} \rangle$, $\lvert z_{2} \rangle$, ..., $\lvert z_{U} \rangle$. The outputs correspond to the expected observations $\langle {z}_{1} \rangle, \langle {z}_{2} \rangle, ..., \langle {z}_{U} \rangle$ arised from the measurement of the Pauli-Z operators. We present the three components in detail next. 

The TPE model was first proposed in \cite{stoudenmire2016supervised}. It aims to convert a classical vector $\textbf{x}$ into a quantum state $\lvert \textbf{x} \rangle$ by setting up a one-to-one mapping as Eq.~(\ref{eq:tpe}). 
\begin{equation}
\begin{split}
\label{eq:tpe}
\vert \textbf{x} \rangle &= \left(\otimes_{i=1}^{U} R_{Y}(\frac{\pi}{2} x_{i}) \right) \vert 0 \rangle^{\otimes U} \\
&= \begin{bmatrix} \cos(\frac{\pi}{2} x_{1}) \\ \sin(\frac{\pi}{2} x_{1}) \end{bmatrix} \otimes \begin{bmatrix} \cos(\frac{\pi}{2} x_{2}) \\ \sin(\frac{\pi}{2} x_{2}) \end{bmatrix} \otimes \cdot\cdot\cdot \otimes \begin{bmatrix} \cos(\frac{\pi}{2} x_{U}) \\ \sin(\frac{\pi}{2} x_{U}) \end{bmatrix},
\end{split}
\end{equation}
where $R_{Y}(\cdot)$ refers to a single-qubit quantum gate rotated across $Y$-axis and each $x_{i}$ is constrained to the domain of $[0, 1]$, which results in a reversely one-to-one conversion between $\textbf{x}$ and $\vert \textbf{x} \rangle$.

Moreover, the PQC is equipped with the CNOT gates for quantum entanglement and learnable quantum gates, i.e., $R_{X}(\alpha_{i})$, $R_{Y}(\beta_{i})$, and $R_{Z}(\gamma_{i})$, where the qubit angles $\alpha_{i}$, $\beta_{i}$, and $\gamma_{i}$ are tuned in the training process. The PQC framework in the green dash square is repeatedly copied to set up a deep model, and the number of the PQC frameworks is called the depth of the VQC. The operation of the measurement outputs the classical expected observations $\vert z_{1}\rangle$, $\vert z_{2}\rangle$, ..., $\vert z_{U}\rangle$ from the quantum output states. The expected outcomes are used to calculate the loss value and the gradient descents~\cite{ruder2016overview}, which are used to update the VQC model parameters by applying the back-propagation algorithm~\cite{werbos1990backpropagation} based on the stochastic gradient descent (SGD) optimizer.

\section{Quantum Natural Gradient Descent}

As shown in Eq. (\ref{eq:gd}), at step $t$, the standard gradient descent minimizes a loss function $\mathcal{L}(\boldsymbol{\theta})$ with respect to the parameters $\boldsymbol{\theta}$ in a Euclidean space. 
\begin{equation}
\label{eq:gd}
\boldsymbol{\theta}_{t+1} = \boldsymbol{\theta}_{t} - \eta \nabla \mathcal{L}(\boldsymbol{\theta}_{t}),
\end{equation}
where $\eta$ is the learning rate. 

The standard gradient descent algorithm conducts each optimization step in a Euclidean geometry on the parameter space. However, since the form of parameterization is not unique, different compositions of parameterizations are likely to distort the distance geometry within the optimization landscape. A better alternative method is to perform the gradient descent in the distribution space, namely natural gradient descent~\cite{amari1998natural}, which is dimension-free and invariant for different parameterization forms. Each optimization step of the natural gradient descent chooses the optimum step size for the update of parameter $\boldsymbol{\theta}_{t}$, regardless of the choice of parameterization. Mathematically, the standard gradient descent is modified as Eq.~\ref{eq:fisher}. 
\begin{equation}
\label{eq:fisher}
\boldsymbol{\theta}_{t+1} = \boldsymbol{\theta}_{t} - \eta F^{-1}\nabla \mathcal{L}(\boldsymbol{\theta}_{t}), 
\end{equation}
$F$ denotes the Fisher information matrix, which acts as a metric tensor that transforms the steepest gradient descent in the Euclidean parameter space to the steepest descent in the distribution space.

Since the standard Euclidean geometry is sub-optimal for the optimization of quantum variational algorithms, a quantum analog has the following form as Eq. (\ref{eq:nw}). 
\begin{equation}
\label{eq:nw}
\boldsymbol{\theta}_{t+1} = \boldsymbol{\theta}_{t} - \eta g^{+}(\boldsymbol{\theta}_{t})\nabla\mathcal{L}(\boldsymbol{\theta}_{t}), 
\end{equation}
where $g^{+}(\boldsymbol{\theta}_{t})$ refers to the pseudo-inverse and is associated with the specific architecture of the quantum circuit. The coefficient $g^{+}(\boldsymbol{\theta}_{t})$ can be calculated using the Fubini-Study metric tensor, which it then reduces to the Fisher information matrix in the classical limit~\cite{mcardle2019variational}.

\section{Quantum Natural Gradient Descent for VQC}

Before employing the QFNGD for a quantum federated learning system, we concentrate on the use of QNGD for a single VQC. For simplicity, we leverage a block-diagonal approximation to the Fubini-Study metric tensor for composing QNGD into the VQC training on the NISQ quantum hardware. 

We set an initial quantum state $\lvert \psi_{0} \rangle$ and a PQC with $L$ layers. For $l\in [L]$, we separately denote $\textbf{W}_{l}$ and $\textbf{V}_{l}(\boldsymbol{\theta}_{l})$ as the unitary matrices associated with non-parameterized quantum gates and parameterized quantum ones, respectively. 

\begin{figure}[htbp]
\centerline{\epsfig{figure=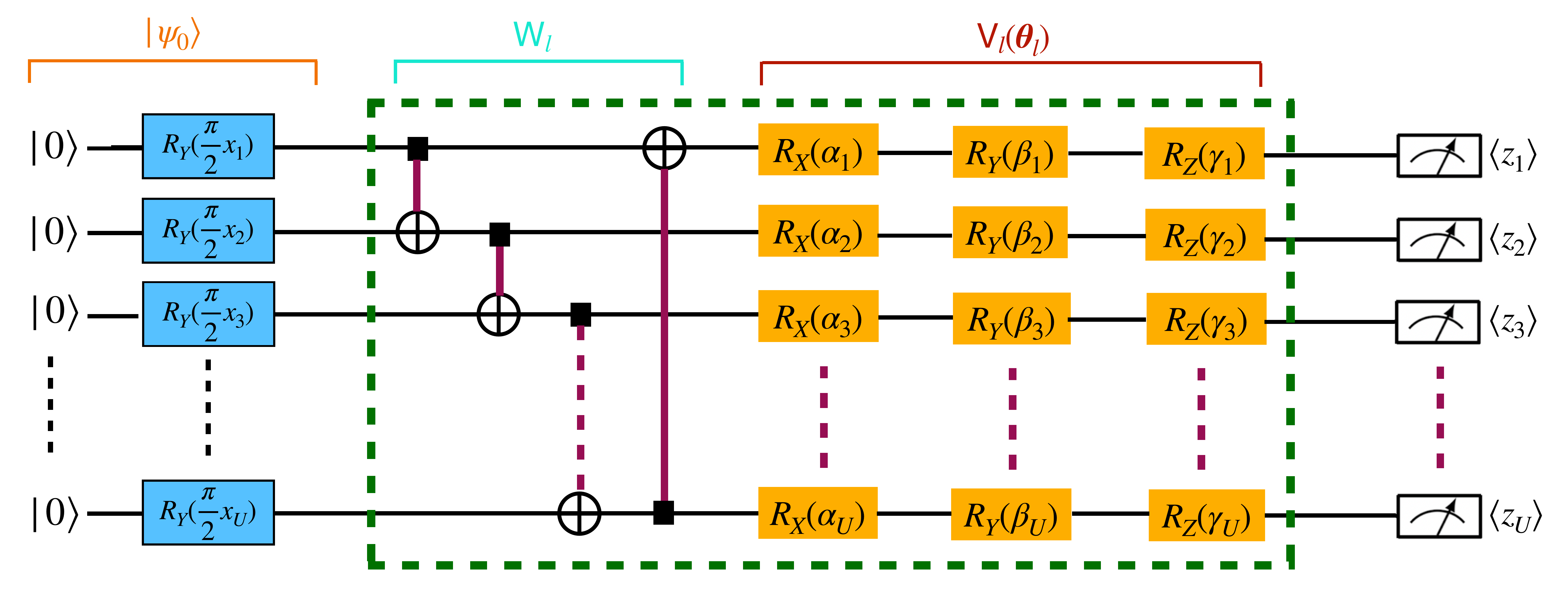, width=80mm}}
\caption{{\it An illustration of unitary matrices associated with the non-parametric and parametric gates. $\forall l\in [L]$, the matrices $\textbf{W}_{l}$ correspond to the non-parametric gates, the matrices $\textbf{V}_{l}(\boldsymbol{\theta}_{l})$ are associated with the parametric ones, and $\lvert \psi_{0} \rangle$ refers to the initial quantum state that is derived from the operation of the TPE.}}
\label{fig:fig2}
\end{figure}

Let's consider a variational quantum circuit as Eq. (\ref{eq:ff}). 
\begin{equation}
\label{eq:ff}
U(\boldsymbol{\boldsymbol{\theta}}) \lvert \psi_{0} \rangle = \textbf{V}_{L}(\boldsymbol{\theta}_{L}) \textbf{W}_{L} \cdot\cdot\cdot \textbf{V}_{l}(\boldsymbol{\theta}_{l}) \textbf{W}_{l} \cdot\cdot\cdot \textbf{V}_{1}(\boldsymbol{\theta}_{1}) \textbf{W}_{1} \lvert \psi_{0} \rangle 
\end{equation}

Furthermore, any unitary quantum parametric gates can be rewritten as $\textbf{V}_{l}(\boldsymbol{\theta}_{l}) = \exp(i \boldsymbol{\theta}_{l} H_{l})$, where $H_{l}$ refers to the Hermitian generator of the gate $\textbf{V}_{L}$. The approximation to the Fubini-Study metric tensor admits that for each parametric layer $l$ in the variational quantum circuit, the $n_{l} \times n_{l}$ block-diagonal submatrix of the Fubini-Study metric tensor $g_{l,i,j}^{+}$ is calculated by Eq. (\ref{eq:app}). 

\begin{equation}
\label{eq:app}
g_{l,i,j}^{+} = \langle \psi_{l} \lvert H_{l}(i) H_{l}(j) \lvert \psi_{l} \rangle - \langle \psi_{l} \lvert H_{l}(i) \lvert \psi_{l} \rangle \langle \psi_{l} \vert H_{l}(j) \vert \psi_{l} \rangle,
\end{equation}
where 
\begin{equation}
\label{eq:part}
\vert \psi_{l} \rangle = \textbf{V}_{l}(\boldsymbol{\theta}_{l})  \textbf{W}_{l} \cdot\cdot\cdot \textbf{V}_{1}(\boldsymbol{\theta}_{1})\textbf{W}_{1} \vert \psi_{0} \rangle. 
\end{equation}

\begin{figure}[htbp]
\centerline{\epsfig{figure=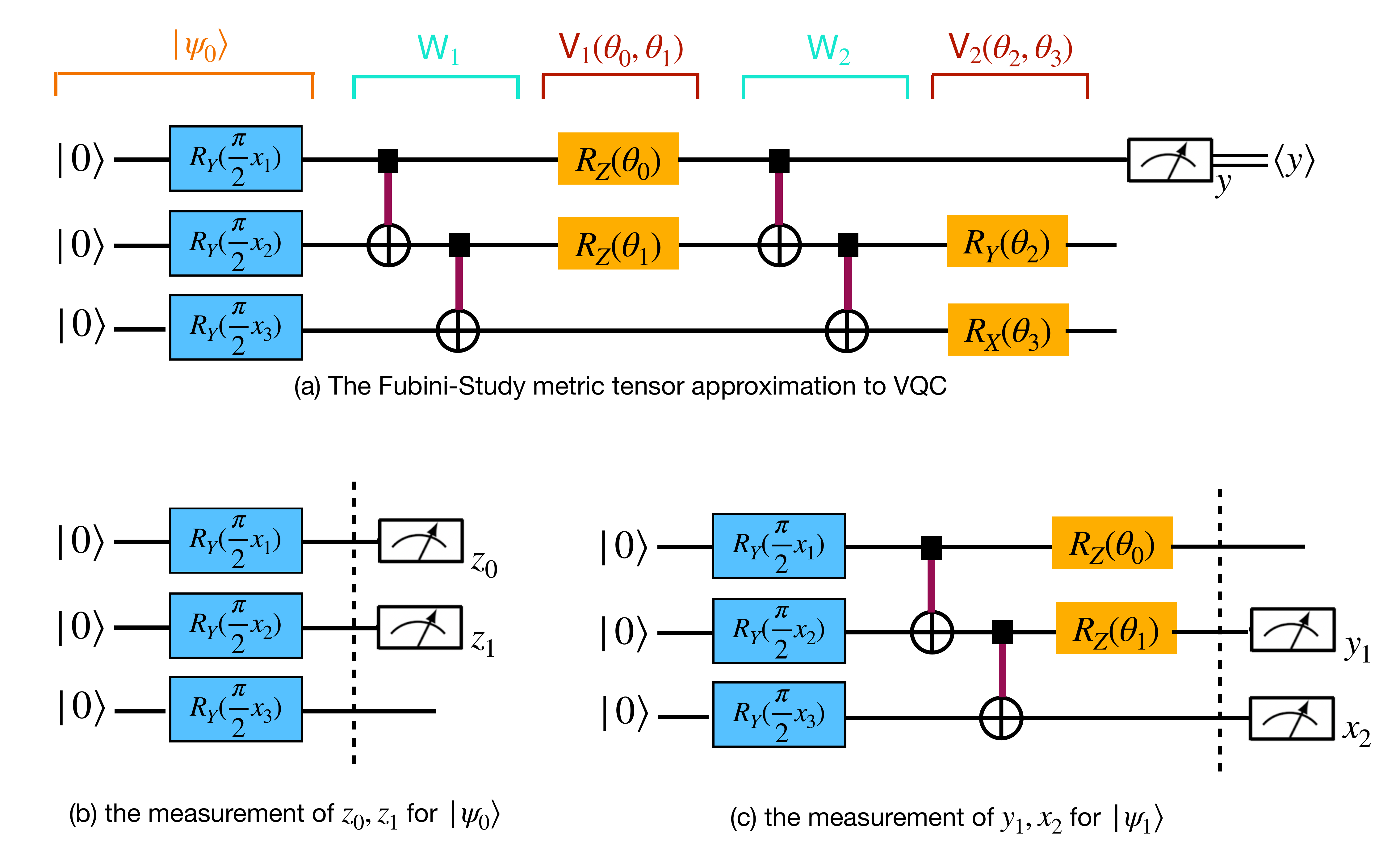, width=80mm}}
\caption{{\it A demonstration of the VQC approximation method based on the Fubini-Study metric tensor: (a) A block-diagonal approximation to VQC based on the Fubini-Study metric tensor; (b) a measurement of $z_{0}, z_{1}$ for $\vert \psi_{0} \rangle$; (c) measurement of $y_{1}, x_{2}$ for $\vert \psi_{1} \rangle$. }}
\label{fig:fig3}
\end{figure}

In Eq. (\ref{eq:part})$, \vert \psi_{l}\rangle$ denotes the quantum state before the application of the parameterized layer $l$. Figure~\ref{fig:fig3} illustrates a simplified version of a VQC, where $\textbf{W}_{1}$ and $\textbf{W}_{2}$ are related to non-parametric gates, and $\textbf{V}_{1}(\theta_{0}, \theta_{1})$ and $\textbf{V}_{2}(\theta_{2}, \theta_{3})$ correspond to the parametric gates with adjustable parameters, respectively. Since there are two layers, each of which owns two free parameters, the block-diagonal approximation is composed of two $2 \times 2$ matrices, $g^{+}_{1}$ and $g^{+}_{2}$, which can be separately expressed as Eq. (\ref{eq:f1}) and (\ref{eq:f2}). 

\begin{equation}
\label{eq:f1}
g^{+}_{1} = \begin{bmatrix}
\langle z_{0}^{2} \rangle - \langle z_{0} \rangle^{2} && \langle z_{0}z_{1} \rangle - \langle z_{0} \rangle \langle z_{1} \rangle  \\
 \langle z_{0}z_{1} \rangle - \langle z_{0} \rangle \langle z_{1} \rangle && \langle z_{1}^{2} \rangle - \langle z_{1} \rangle^{2}
\end{bmatrix},
\end{equation}
and 
\begin{equation}
\label{eq:f2}
g^{+}_{2} = \begin{bmatrix}
\langle y_{1}^{2} \rangle - \langle y_{1} \rangle^{2} && \langle y_{1}x_{2} \rangle - \langle y_{1} \rangle \langle x_{2} \rangle  \\
 \langle y_{1}x_{2} \rangle - \langle y_{1} \rangle \langle x_{2} \rangle && \langle x_{2}^{2} \rangle - \langle x_{2} \rangle^{2}
\end{bmatrix}.
\end{equation}

The elements of $g_{1}^{+}$ and $g_{2}^{+}$ compose $g^{+}(\boldsymbol{\theta})$ as Eq. (\ref{eq:fff}). 
\begin{equation}
\label{eq:fff}
g^{+}(\boldsymbol{\theta}) = \begin{bmatrix}
g_{1}^{+} & 0 \\
0 & g_{2}^{+}
\end{bmatrix}. 
\end{equation}

Then, we employ Eq.~(\ref{eq:nw}) to update the VQC parameter $\boldsymbol{\theta}$. 

\section{Federated Quantum Natural Gradient Descent}

\begin{algorithm}[tb]
   \caption{Iterative Approximation}
   \label{alg:iterative}
 1. Given the dataset $S = S_{1} \cup S_{2} \cup \cdot\cdot\cdot \cup S_{K}$. \\
 2. Initialize global parameter $\bar{\boldsymbol{\theta}}$ and broadcast it to participants $\boldsymbol{\theta}_{0}^{(k)}$.  \\
 3. Assign each participant with the subset $S_{k}$. \\
 4. For each global model update at epoch $t=[T]$ do \\
 5. \hspace{4mm} For each participant $k \in [K]$ \textbf{in parallel} do \\
 6. \hspace{7mm} Attain $g^{+}(\boldsymbol{\theta}^{(k)}_{t}; S_{k}) \nabla \mathcal{L}(\boldsymbol{\theta}_{t}^{(k)})$ for the $k^{th}$ VQC.  \\
 7. \hspace{7mm} Send gradient $g^{+}(\boldsymbol{\theta}^{(k)}_{t}) \nabla \mathcal{L}(\boldsymbol{\theta}_{t}^{(k)}; S_{k})$ to the coordinator. \\
 8. \hspace{4mm} End for \\
 9. \hspace{3.1mm} The coordinator aggregates the received gradients.  \\
 10.\hspace{3.0mm} The coordinator updates the global model by Eq. (\ref{eq:aupdate}). \\
 11. \hspace{3mm}Broadcast the updated global $\bar{\boldsymbol{\theta}}_{t+1}$ to all participants. \\
 12. End for
\end{algorithm}

A QFL system can be built by setting up VQC models in an FL manner, given the dataset $S$ composed of subsets $S_{1}, S_{2}, ..., S_{K}$, the objective of QFL can be formulated as:
\begin{equation}
\min\limits_{\boldsymbol{\theta}} \sum\limits_{k=1}^{K} w_{k} g^{+}_{k}(\boldsymbol{\theta})\mathcal{L}(\boldsymbol{\theta}; S_{k}),
\end{equation}
where $w_{k}$ refers to the coefficient assigned to the $k$-th gradient participant, and each $w_{k}$ can be estimated as:
\begin{equation}
w_{k} = \frac{|S_{k}|}{|S|} = \frac{|S_{k}|}{\sum_{k=1}^{K}|S_{k}|}. 
\end{equation}

The QNGD algorithm is applied for each VQC and the uploaded gradients of all VQCs are aggregated to update the model parameters of the global VQC. The FQNGD can be mathematically summarized as:

\begin{equation}
\label{eq:aupdate}
\bar{\boldsymbol{\theta}}_{t+1} = \bar{\boldsymbol{\theta}}_{t} - \eta \sum\limits_{k=1}^{K} \frac{|S_{k}|}{|S|} g_{k}^{+}(\boldsymbol{\theta}^{(k)}_{t}) \nabla\mathcal{L}(\boldsymbol{\theta}^{(k)}_{t}; S_{k}),  
\end{equation}
where $\bar{\boldsymbol{\theta}}_{t}$ and $\boldsymbol{\theta}_{t}^{(k)}$ separately correspond to the model parameters of the global VQC and the $k$-th VQC model at epoch $t$, and $N_{k}$ represents the amount of training data stored in the participant $k$, and the sum of $K$ participants' data is equivalent to $N$.

Compared with the SGD counterparts used for QFL, the FQNGD algorithm admits adaptive learning rates for the gradients such that the convergence rate could be accelerated according to the VQC model status.

\subsubsection{Empirical Results}

To demonstrate the FQNGD algorithm for QFL, we perform the binary and ternary classification tasks on the standard MNIST dataset~\cite{deng2012mnist}, with digits $\{2, 5\}$ for the binary task and $\{1, 3, 7\}$ for the ternary one. There are $11379$ training data and $1924$ test data for the binary classification, and $19138$ training data and $3173$ test data are assigned for the ternary classification. As for the setup of QFL in our experiments, the QFL system consists of $6$ identically local VQC participants, each of which owns the same amount of training data. The test data are stored in the global part and are used to evaluate the classification performance. 

\begin{figure}[htbp]
\centerline{\epsfig{figure=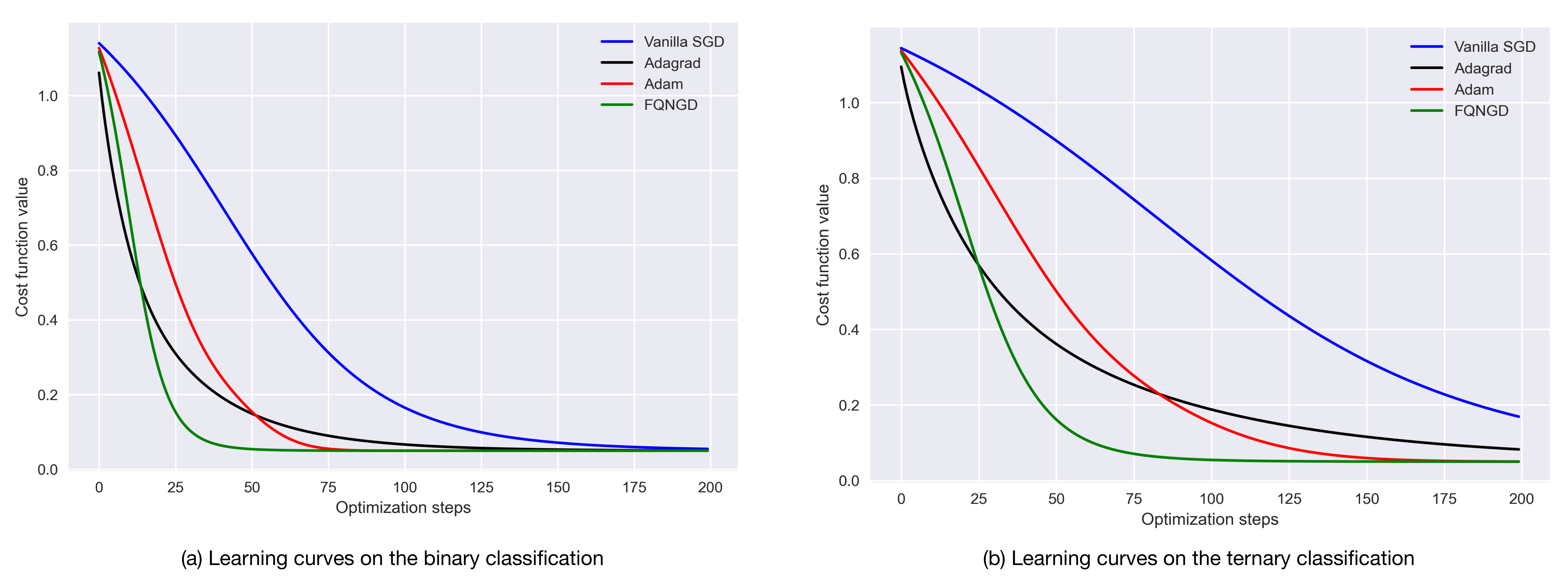, width=95mm}}
\caption{{\it Simulation results of binary and ternary classifications on the training set of the MNIST database. (a) The learning curves of various optimization methods for the binary classification; (b) the learning curves of various optimization methods for the ternary classification.}}
\label{fig:res}
\end{figure}

\begin{table}[tpbh]\footnotesize
\center
\renewcommand{\arraystretch}{1.3}
\caption{The simulation results of a binary classification.}
\begin{tabular}{|c||c|c|c|c|}
\hline
Methods     		& Vanilla SGD		&   Adagrad     	& Adam	 & FQNGD   \\
\hline
Accuracy			&	98.48		&	98.81	&	98.87	&	\textbf{99.32}	\\
\hline
\end{tabular}
\label{tab:tab1}
\end{table}

\begin{table}[tpbh]\footnotesize
\center
\renewcommand{\arraystretch}{1.3}
\caption{The simulation results of a ternary classification.}
\begin{tabular}{|c||c|c|c|c|}
\hline
Methods     		& Vanilla SGD		&   Adagrad     	& Adam	& FQNGD   \\
\hline
Accuracy			& 97.86		& 98.63	&98.71	&	\textbf{99.12}	\\
\hline
\end{tabular}
\label{tab:tab2}
\end{table}

We compare our proposed FQNGD algorithm with other three optimizers: the naive SGD optimizer, the Adagrad optimizer~\cite{lydia2019adagrad}, and the Adam optimizer~\cite{kingma2014adam}. The Adagrad optimizer is a gradient descent optimizer with a past-gradient-dependent learning rate in each dimension. The Adam optimizer refers to the gradient descent method with an adaptive learning rate as well as adaptive first and second moments.

As shown in Figure~\ref{fig:res}, our simulation results suggest that our proposed FQNGD method is capable of achieving the fastest convergence rate among the optimization approaches. It means that the FQNGD method can reduce the communication overhead cost and maintain the baseline performance of binary and ternary classifications on the MNIST dataset. Moreover, we evaluate the QFL performance in terms of classification accuracy. The FQNGD method outperforms the other counterparts with the highest accuracy values. In particular, the FQNGD is designed for the VQC model and can attain better empirical results than the Adam and Adagrad methods with adaptive learning rates over epochs. 

\section{Conclusion and Future Work}
This work focuses on the design of the FQNGD algorithm for the QFL system in which multiple local VQC models are applied. The FQNGD is derived from training a single VQC based on QNGD, which relies on the block-diagonal approximation of the Fubini-Study metric tensor to the VQC architecture. We put forth the FQNGD method to train the QFL system. Compared with other SGD methods such as Adagrad and Adam optimizers, our experiments of the classification tasks on the MNIST dataset demonstrate that the FQNGD method attains better empirical results than other SGD methods, while the FQNGD exhibits a faster convergence rate than the others, which implies that our FQNGD method suggests that it is capable of reducing the communication cost and can maintain the baseline empirical results. 

Although this work focuses on the optimization methods for the QFL system, the decentralized deployment of a high-performance QFL system for adapting to the large-scale dataset is left for our future investigation. In particular, it is essential to consider how to defend against malicious attacks from adversaries and also boost the robustness and integrity of the shared information among local participants. Besides, the deployment of other quantum neural networks like quantum convolutional neural networks (QCNN)~\cite{cong2019quantum} are worth further attempts to compose a QFL system. 

\clearpage
% References should be produced using the BibTeX program from suitable
% BibTeX files (here: strings, refs, manuals). The IEEEbib.bst bibliography
% style file from IEEE produces an unsorted bibliography list.
% -------------------------------------------------------------------------
\footnotesize
\bibliographystyle{IEEEbib}
\bibliography{refs}

\end{document}